\documentclass[aps,prb,preprint,showpacs,showkeys,groupedaddress]{revtex4}
\usepackage{graphicx}
\usepackage{amssymb,amsmath}

\begin{document}

\title{ZnO:Co Diluted Magnetic Semiconductor or Hybrid Nanostructure for Spintronics?}

\author{F. Golmar}
 \affiliation{Lab. de Ablación L\'{a}ser, Dto. F\'{i}sica, Facultad de Ingeniería, Universidad de Buenos Aires, Argentina.}

\author{M. Villafuerte}
 \affiliation{Laboratorio de F\'{i}sica del S\'{o}lido,  Dpto. de
F\'{i}sica, Facultad de Ciencias Exactas y Tecnolog\'{i}a,
Universidad Nacional de Tucum\'{a}n. CONICET, Argentina.}

\author{A. Mudarra Navarro}
 \affiliation{Dto. F\'{i}sica-IFLP, Facultad de Ciencias Exactas, Universidad Nacional de La Plata. CONICET. Argentina}

\author{C. E. Rodr\'{i}guez Torres}
 \affiliation{Dto. Física-IFLP, Facultad de Ciencias Exactas, Universidad Nacional de La Plata. CONICET. Argentina}

\author{J. Barzola-Quiquia}
\affiliation{Division of Superconductivity and Magnetism, Institut fur Experimentelle Physik II, Universitat, Leipzig, Germany}

\author{S. P. Heluani}
\affiliation{Laboratorio de F\'{i}sica del S\'{o}lido,  Dpto. de
F\'{i}sica, Facultad de Ciencias Exactas y Tecnolog\'{i}a,
Universidad Nacional de Tucum\'{a}n, Argentina.}

\author{P. Esquinazi}
\affiliation{Division of Superconductivity and Magnetism, Institut fur Experimentelle Physik II, Universitat, Leipzig, Germany}

\date{\today}

\begin{abstract}
 We have studied the influence of intrinsic and extrinsic defects in the magnetic and electrical transport properties of Co-doped ZnO thin films.  X ray absorption measurements show that Co substitute Zn in the ZnO structure and it is in the 2+ oxidation state. Magnetization (M) measurements show that doped samples are mainly paramagnetic. From M vs. H loops measured at 5 K we found that the values of the orbital L and spin S numbers are between 1 and 1.3 for L  and S = 3/2, in agreement with the representative values for isolated Co 2+. The obtained negative values of the Curie-Weiss temperatures indicate the existence of antiferromagnetic interactions between transition metal atoms.

\end{abstract}



\maketitle

\section{Introduction}
Transition metal atoms introduced into the cationic sites of semiconducting host lattices have recently attracted increasing attention because of their potential use in spintronic devices. Since the theoretical prediction of ferromagnetism in Co-doped ZnO, this material is being intensively studied \cite{Dietl:Sci00}.  However, contradicting experimental and theoretical results, concerning magnetic properties, have been reported\cite{Ueda:APL01, Kim:JAP02, Venkatesan:PRL04, Chang:JPD07, Ney:PRL08}. Some groups observe room temperature ferromagnetism, while others report no ferromagnetism at room temperature or the observed ferromagnetism comes from metallic cobalt clusters. Also, the mechanism of exchange coupling induced by defect states, such as O- and Zn-vacancies, Zn- and N-interstitial or due to an inhomogeneous dopant distribution in the ferromagnetism remains still unclear.

An exchange coupling mechanism has been proposed by Dietl {\it et al.} \cite{Dietl:Sci00} where the wave function of an unpaired electron in the valence band overlap the wave function of the transition metal ion having an opposite spin. The impurity band can interact with local magnetic moments through the formation of bound magnetic polaron (BMP)\cite{Dietl:Sci00}. The $Co$ ions are often found in the $Co^{+}$ states in $ZnO$, and it is necessary to introduce defects with a spin to provide the magnetic state necessary for the exchange coupling \cite{Paterson:PRB06}.  In a recent work \cite{Dietl:PRB07} Dietl {\it et al.} describe the origin of the ferromagnetic signal in Co doped $ZnO:Al$ in terms a spinodal decomposition of a Co-rich antiferromagnetic nanocrystals embedded in a Co poor ZnO:Co host material. The authors describe the ferromagnetism as a consequence of the uncompensated spins at the antiferromagnetic nanocrystals surface.   Recently, Coey {\it et al.} \cite{Coey:Nat05} argument that traditional super-exchange and double exchange interaction cannot produce long range magnetic order at concentration of magnetic cations of a few per cent, and they proposed another model for ferromagnetism in DMS materials based on a spin-split donor impurity band. The ferromagnetism exchange in this model is mediated by the defect states that form bound magnetic polaron generating an impurity band which interact with local magnetic moments. To obtain a high Tc, a fraction of polaronic charge must be delocalized (or hybridized). This description of ferromagnetic mechanism opens the possibility to control de magnetic behavior of DMS by injection of carrier using photons or electric field. The latter description suggest us to start a study on electric and magneto-transport mechanisms, and the effect of the interaction between defects and dopants on these mechanisms.

In this work, 10\% and 15\% Cobalt doped ZnO films were grown on SiO2/Si and sapphire substrates in N2 or O2 atmospheres by pulsed laser deposition (PLD). In order to understand how the micro or nano-structural inhomogeneity can affect the magnetic and electrical properties of these samples, these properties were investigated by measuring the in-plane magnetization versus magnetic field and also the electrical and electromagnetic transport as a function of temperature between 5K and room temperature. These results were correlated with structural and compositional studies. All the samples of the series doped with Co showed a paramagnetic behavior. The pure samples were diamagnetic, independently of the growth atmosphere. A giant positive magnetoresistance was estimated using IV curves taken at constant temperatures in zero magnetic field and in 10 Tesla. Structural and magnetic characterizations suggest us to consider the system ZnO:Co as an hybrid nanostructure.

\section{Experimental}
Two series of ZnO films of different thickness were prepared by PLD. A Nd:YAG laser beam, operating at 10 Hz, was focused on a rotating target of ZnO:Co with the nominal composition, to yield an energy density of 3.4 J/cm2. In both series, 10\% and 15\% Cobalt doped ZnO films were grown on SiO2/Si and c-sapphire substrates in N2 or O2 atmospheres. The substrates were held at a temperature of 673K. For the first serie a laser wavelenght of 266 nm was used and thin films with thicknesses between 50 and 100nm were deposited, and for the second serie thicker films with thicknesses from 800 to 900 nm were deposited using a laser wavelenght of 355 nm.
X-ray diffraction (XRD) and scanning electron microscopy with Energy Dispersive Spectroscopy (EDS) was employed for structural characterizations. Magnetization measurements were performed using a superconducting quantum interference device (SQUID) magnetometer at temperatures
from 5 to 300 K. For electrical transport measurements a symmetric device was fabricated by depositing two coplanar Al disk-shaped
contacts (diameter = 0.75 mm) by Ar-sputtering on the top surface of the
samples. Leads were soldered on both pair of contacts using In. We employed a two-probe scheme in the setup. In the two-probe method, a fixed voltage was applied across the sample, and the current flow was monitored by measuring the voltage drop across a metal film resistor. This scheme turns out to be more convenient because it allows the stabilization of the dc voltage at sufficiently high level, and that fixed value is not affected by dramatic changes of resistivity at low temperatures. Leakage current values were checked to be always below the current values measured in this work.  The I-V  curves were constructed using voltage pulses of increasing amplitude of about one second time width, followed by a short time (half a second) without any applied voltage.

\section{Results}

X rays diffraction pattern (XRD) corresponded to the wurtzite structure for all films showing only the (002) and (004) wurtzite reflection.  Figure 1a shows as an illustration the patterns corresponding to a 15\% Co-doped ZnO thick film of the serie growth on sapphire compared with a pure sample growth in identical conditions. The double peaks are observed only in the doped and thicker films. Figure 1b shows a magnification of the double peak at $2\theta=34.08^{0}$. This result could be attributed  to a (002) reflection from a composition modulated alloy of two wurtzite systems with different lattice parameter due to dissimilar Co distribution. It was already reported the coincidence of the (002) ZnO and $w-CoO$ reflection \cite{Alaria:JPD08}, then it is difficult to resolve the presence of both phases.

\begin{figure}
 \includegraphics{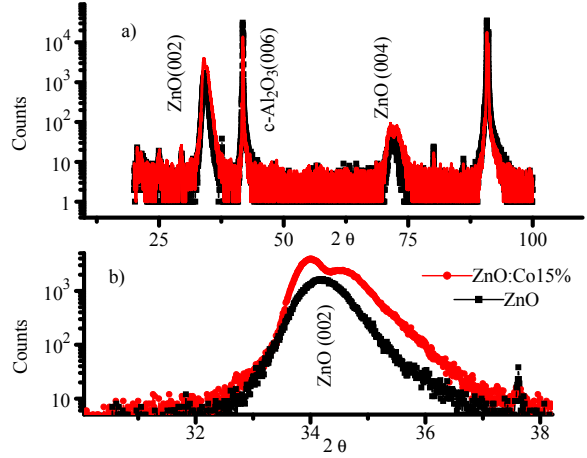}
 \caption{\small{\it{XRD spectra of films of ZnO deposited on sapphire prepared by PLD, black line: pure $ZnO$, red line: $Zn_{0.85}OCo_{015}$ a)Full spectra; b)zoom of the (002) peaks}}}
     \end{figure}

The XRD patterns of thinner films, all growths on SiO2/Si substrate show a polycrystalline lightly oriented structure. Although no double peaks appear, XRD revealed the segregation of Co particles.

 \begin{figure}
 \includegraphics{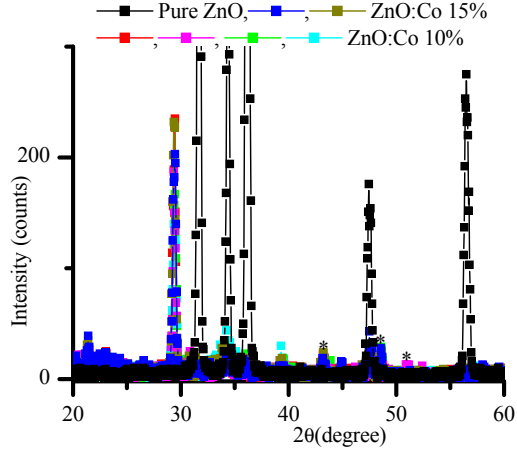}
 \caption{\small{\it{XRD spectra of thinner films of ZnO deposited on SiO2/Si, the size of samples are between 50 and 100nm, asterisks indicate the Co impurity}}}
     \end{figure}

Figures 3a and 3b show a scanning electron microscopy with Energy Dispersive Spectroscopy (EDS) mapping showing the Zn and Co distribution in  $Zn_{0.85}OCo_{015}$ film grown over sapphire substrates in N2 atmosphere (the same doped film as figure 1). $Co \ K_{\alpha}$  and $Zn \ L $ X-ray lines were used for the Co and Zn EDS mapping respectively.  Although the Zn distribution seems uniform (similar distribution is obtained for the oxygen, not shown) the segregation observed in the Co EDS mapping seem like a typical spinodal decomposition. From XRD pattern an average relation of 60/1000 nm between the sizes of the particles of different composition it was estimated using the Scherrer relation, $D= 0.9. \lambda /\Omega cos(\theta)$, where $\lambda$ is the x ray wavelength, $\Omega$ is the full with at half maximum at 002 peak and $\theta$ is the diffraction angle of the XRD spectra.  A mean sizes of particles richer in $Co$  were estimated as 60 nm.

 Considering that no nanoparticles of CO, of any other phase, of size below 10nm could be resolved using X rays data and, in addition, thinner sample EDS mapping shows similar segregation than thicker samples, our results suggest us to consider the system ZnO:Co as an hybrid nanostructure.

X ray absorption spectroscopy, figure 3c, showed that Co ions are in 2+ state, substituting Zn in the ZnO structure.

\begin{figure}
 \includegraphics{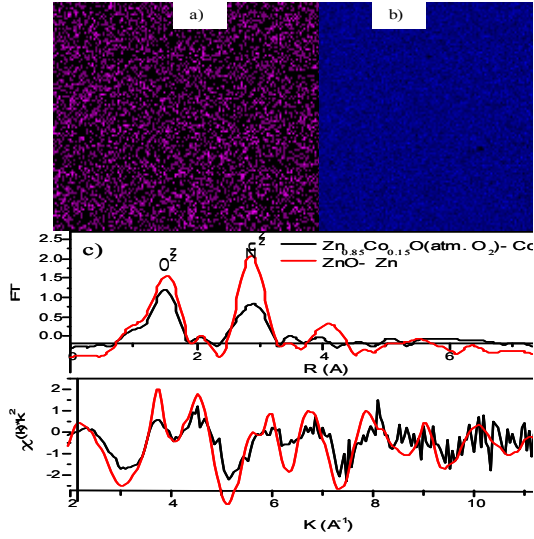}
 \caption{\small{\it{Energy Dispersive Spectroscopy mapping showing the Zn and Co distribution in a same sample using $Co / K_{\alpha}$ in a)  and $Zn / L $ x-ray lines in b). c) XAS spectra of $Zn_{0.85}OCo_{0.15}$ films.}}}
     \end{figure}

IV curves of ZnO thin film doped with 10\% Co shows in Figure 4. The IV curves were taken at constant temperatures in zero magnetic field and in 10 Tesla. Using these data we can evaluate the magnetoresistance as MR = (R(H) - R(0))/R(0) , the results show that the MR is positive and increases when the temperature is decreased. At 50K the MR reach 30\%. This large positive MR already founded in other DMS \cite{Andrearczyk:PRB05} and it was explained considering the spin-orbit scattering \cite{Sawicki:PRL86}.

\begin{figure}
 \includegraphics{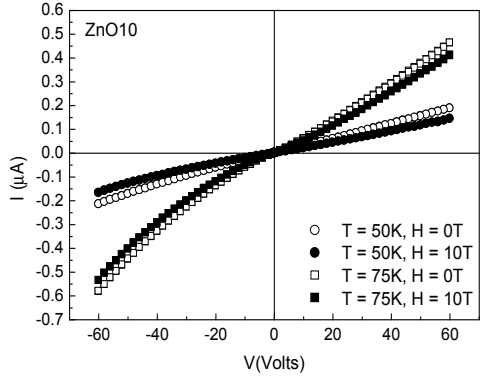}
 \caption{\small{\it{IV curves of ZnO thin film doped with 10\% Co in zero field and in 10T)}}}
     \end{figure}

Figure 5a shows room temperature SQUID measurements of Co doped thinner samples. A zoom near the origin is shown in figure 5b, we can see that the values of the coercivity are similar for all samples. The thinner samples have similar values of saturation moment with the exception of the 50nm sample. The M-H loop for the non doped sample is also shown for comparison. In Figure 6, a SQUID loop for representative thicker samples showing the paramagnetic behavior at 5K for most of them. The hysteresis  observed in some thicker samples is similar in magnitude than thinner sample, suggesting that the origin of ferromagnetic signal it is as consequence of interfaces or impurities.

SQUID loops for thinner and thicker data can be fitted well with the Brillouin function for $L=1$ and $S=3/2$ and $L=1.3$ and $S=3/2$, respectively. The fitting variable was the magnetic saturation, yielding values between $\mu=4.8\mu_{B}/Co$ and $\mu=5.0\mu_{B}/Co$. This result agree with the value expected for Co 2+. By taking into account the sample volume and composition the magnetization expected with the moments calculate by Brillouin are one order of magnitude larger than the SQUID measurements. The minor values in the measured magnetization could be explained by the contribution of an antiferromagnetic exchange coupling.

 \begin{figure}
\includegraphics{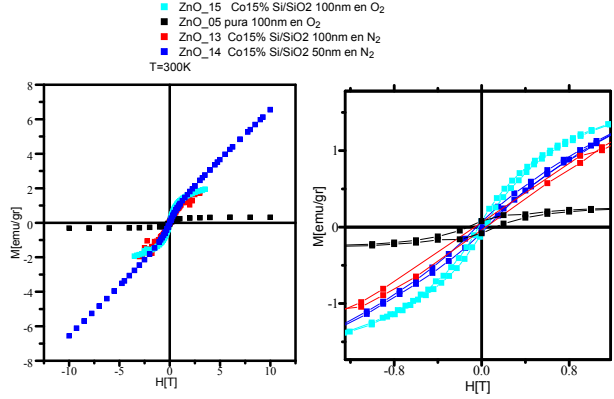}
  \caption{\small{\it{M vs. H loops taken at 300 K for $Zn_{0.85}OCo_{0.15}$ and $ZnO$ films in $O_{2}$, $N_{2}$.}}}
     \end{figure}

The negative values of the Curie Weiss temperature obtained from the extrapolation of the high temperature range of the inverse moment vs. temperature, $(-280\pm 7k)$ for a thicker film and $(-166.6\pm3.8)$K for a thinner film, indicate the existence of antiferromagnetic interactions, inset of figure 6. This behavior may be ascribed to the antiferromagnetism of CoO clusters. Antiferromagnetic exchange of Co clusters in single crystalline ZnO was recently theoretical and experimentally  studied  by Sati {\it{et al}} \cite{Sati:PRL07}.

 \begin{figure}
\includegraphics{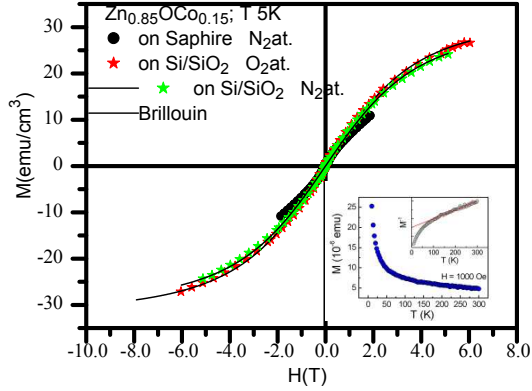}
  \caption{\small{\it{M vs. H loops taken at 5 K for $Zn_{0.85}OCo_{0.15}$ and $ZnO$ films in $O_{2}$, $N_{2}$. Inset: M vs. T curve for a $Zn_{0.85}OCo_{0.15}$ film.}}}
     \end{figure}

  \section{Conclusion}
In this letter we try to explain the origin of the controversial results reported in the literature in polycrystalline Co-doped ZnO films below the limited solubility of Co in ZnO. Structural, magnetic and Magneto-transport measurements reported here suggest that the formation of Co reach clusters, as CoO in its wurzite or cubic structure, is the most probable cause of the magnetic interaction in Co:ZnO. We found paramagnetic behavior in all doped samples with a small ferromagnetic contribution in a few samples which could be attributed to interfaces or small amount of impurities. The paramagnetic behavior was observed at room temperature down to 5K. By evaluate our magnetic measurements we reach at the conclusion that the Co reach clusters has an antiferromagnetic exchange interaction.

In summary, here we present evidence of segregation in Co doped ZnO and its consequence in magnetic measurements. Considering the system ZnO:Co as an hybrid nanostructure, the  possibility to describe the magnetic interaction using the DMS models recently report \cite{Dietl:PRB07,Coey:Nat05} it is difficult.

\begin{acknowledgments}
This work has been supported by CIUNT under Grants
26/E439 by ANPCyT-PICTR 20770 and 35682, by PROALAR-DAAD DA0805.
\end{acknowledgments}

\end{document}